\documentclass[twocolumn,preprintnumbers,amsmath,amssymb]{revtex4}
\usepackage{graphicx}
\usepackage{dcolumn}
\usepackage{bm}

\newcommand{\be}{\begin{equation}}
\newcommand{\bel}[1]{\begin{equation}\label{#1}}
\newcommand{\ee}{\end{equation}}
\newcommand{\bea}{\begin{eqnarray}}
\newcommand{\ba}{\begin{array}}
\newcommand{\eea}{\end{eqnarray}}
\newcommand{\ea}{\end{array}}

\begin{document}

\title{\bf Asymmetric Exclusion Processes with Disorder: Effect of Correlations }
\author{M. Ebrahim Foulaadvand ~\footnote{Corresponding author: foolad@iasbs.ac.ir}}
\affiliation{Department of Nano-Science,
 Institute for Studies in Theoretical Physics and Mathematics (IPM),
P.O. Box 19395- 5531,Tehran, Iran and Department of Physics, Zanjan
University, P.O. Box 45195-313, Zanjan,  Iran.}

\author{Anatoly B. Kolomeisky}
\affiliation{ Department of Chemistry, Rice University, Houston, TX
77005 USA. }

\author{Hamid Teymouri}
\affiliation{Department of Physics, Zanjan University, P.O. Box
45195-313, Zanjan,  Iran.}

\date{\today}
\begin{abstract}

Multi-particle dynamics in one-dimensional asymmetric exclusion
processes with disorder is investigated theoretically by
computational and analytical methods. It is argued that the
general phase diagram consists of three non-equilibrium phases
that are determined by the dynamic behavior at the entrance, at
the exit and at the slowest defect bond in the bulk of the
system. Specifically, we consider dynamics of asymmetric
exclusion process with two identical defect bonds as a function
of distance between them. Two  approximate theoretical methods,
that treat the system as a sequence of segments with exact
description of dynamics inside the segments and neglect
correlations between them, are presented. In addition, a numerical
iterative procedure for calculating dynamic properties of
asymmetric exclusion systems is developed. Our theoretical
predictions are compared with extensive Monte Carlo computer
simulations. It is shown that correlations play an important role
in the particle dynamics. When two defect bonds are far away from
each other the strongest correlations are found at these bonds.
However, bringing defect bonds closer leads to the shift of
correlations to the region between them. Our analysis indicates
that it is possible to develop a successful theoretical
description of asymmetric exclusion processes with disorder by
properly taking into account the correlations.

\end{abstract}

\maketitle
\section{{Introduction}}

In recent years a significant attention has been devoted to
investigation of low-dimensional asymmetric simple exclusion
processes (ASEPs)
\cite{derrida98,zia95,schutz,chowdhury00,derrida93}. They play  a
critical role for understanding  fundamental properties of
non-equilibrium phenomena in Chemistry, Physics and Biology.
ASEPs have been widely utilized for description of traffic
phenomena \cite{chowdhury00}, kinetics of biopolymerization
\cite{macdonald68}, protein synthesis
\cite{shaw03,chou04,shaw04,dong07}, and biological transport of
motor proteins \cite{lipowsky01,parmeggiani03}. The advantage of
using asymmetric exclusion processes for studying mechanisms of
non-equilibrium phenomena is due to the fact that some
homogeneous versions of ASEPs can be solved exactly via
matrix-product approach and related methods
\cite{derrida98,derrida93,blythe07}. In addition, understanding
of processes in ASEPs can be achieved by utilizing a
phenomenological domain wall approach \cite{DW}. In order to have
a more realistic description of different non-equilibrium
phenomena ASEPs with inhomogeneous distribution of rates are
required. However, there is a limited number of  studies dealing
with ASEPs with disorder in the transition rates at sites (static impurities)
\cite{chou04,shaw04,janowsky92,schutz93,schutz93PRE,janowsky94,kolomeisky98,tripathy97,barma98,kolwanker,ha,stinchcombe,derrida04,juhasz05,barma06,pierobon06,foulaadvand07,foulad07,dongPRE,greulich1,greulich2}
and with disorder associated to particles's hopping rates (moving impurities) \cite{speers,mallick,evans,kim,fouladvand99}.
In this case exact solutions are not obtained, and extensive
Monte Carlo computer simulations and approximate theories are
utilized in order to understand particle dynamics. Disorder has a
strong effect on the behavior of ASEPs. Even a single defect bond
far away from the boundaries lead to dramatic effects in the
stationary properties both in closed
\cite{janowsky92,janowsky94}  and open boundary conditions
\cite{kolomeisky98}.  It was shown recently that the dynamics of
ASEPs is also influenced by several defects that are close to
each other \cite{chou04,dong07,greulich1}, although the mechanism
of this phenomenon is not well understood. This interaction
between  defects is important for understanding several
biological transport phenomena \cite{chou04,dong07}.  Recently
the particular case of two defects has been extensively
investigated by Monte Carlo simulations \cite{dong07}. It has been
shown that the system current exhibits a notable dependence on the
distance between defects with equal hopping rates. Moreover, it
was found that the density profile is linear between defects
which marks the existence of wandering shock between defects
\cite{dong07}. The case of two defective sites with equal rate
has been generalized to include extended objects \cite{dongPRE}.
Theoretical efforts to analyze ASEPs with disorder have been
mostly directed to the cases with a single or few defects
\cite{kolomeisky98,chou04,greulich1}. In Ref. \cite{kolomeisky98}
ASEP with open boundaries  and with a local inhomogeneity in
the bulk  has been investigated by arguing that the defect bond
divides the system into two coupled homogeneous ASEPs. This
theoretical approach can be called a {\it defect mean-field}
(DMF) because the mean-field assumptions are made only at the
position of  local inhomogeneity.  Although a good agreement with
computer simulations has been found, there were significant
deviations in statistical properties of the phase with the
maximal current that was attributed to the neglect of
correlations at the defect bond in the proposed theory
\cite{kolomeisky98}. A related approach called {\it interacting
subsystem approximation} (ISA) has been proposed in Ref.
\cite{greulich1} for ASEPs with a single defect or several
consecutive defects (bottleneck). Here it was suggested that due
to the defect bonds there are 3 segments in the system: two
homogeneous ASEPs are coupled by a segment that includes all
sites that surround defect bonds. Explicit results have been
used  inside the segments, and  mean-field assumptions have been
utilized for particle dynamics between the segments. A better
agreement with Monte Carlo computer simulations has been found,
and the method was also successfully applied to describe
interactions of defects with boundaries. It was argued that ISA
can be used for analyzing properties of general ASEPs with
disorder \cite{greulich1,greulich2}. However, ISA has not been
applied for the systems with 2 defects at finite distances from
each other, and because of this observation it is difficult to
apply ISA for understanding mechanisms of more complex
inhomogeneous asymmetric exclusion processes. A slightly
different method of calculations has been proposed by Chou and
Lakatos \cite{chou04}, who applied a {\it finite segment
mean-field theory} (FSMFT). According to this approach, the
segment of finite length $n$ that covers the defect and
surrounding sites is considered, and its dynamics is fully
described by solving explicitly for eigenvectors of the
corresponding transition rate matrix. The segment is then coupled
in the mean-field fashion to the rest of the system. However,
this approach becomes numerically quite involved for cluster
sizes  larger than $\approx$20, and it also limits its
applicability. Different studies of asymmetric exclusion
processes with disorder point out to importance of correlations
in the system. It is reasonable to expect that correlations are
stronger near the slow defect sites. However, it is not clear how
far from the local inhomogeneity and how fast these correlations
decay. In addition, it is also unclear  how correlations from two
close defects affect each other. The goal of this paper is to
investigate the role of correlations in dynamics of ASEPs with
disorder. By analyzing several analytical approaches in
combination with extensive Monte Carlo computer simulations it
will be shown that a successful description of disordered driven
diffusive systems can be achieved by properly accounting for
correlations near the defect bonds.

\section{Model and Theoretical Description}

We investigate a totally asymmetric simple exclusion processes
with disorder. In the one-dimensional lattice
the particle at the site $i$ can jump forward with the rate
$p_{i}$ if the next site $i+1$ is unoccupied, otherwise it stays
at the same place. The particle can enter the system with the
rate $\alpha$ if the site is empty, and it can also exit the
lattice with the rate $\beta$. When all $p_{i}=1$ we have a
homogeneous ASEP for which dynamic properties are known
explicitly from exact solutions \cite{derrida98,schutz}.
ASEPs with disorder correspond to the situation when there is
inhomogeneities in the transition rates, and $p_{i}$ are drawn
from arbitrary distributions. Numerous theoretical and
computational studies indicate that in the limit of large times
the dynamics in the system can be determined by comparing
entrance rate, exit rates and the transition rate at the slowest
defect bond \cite{kolomeisky98,greulich2}. This observation has a
significant consequence for properties of ASEPs with disorder,
yielding a generic  phase diagram with 3 phases.  When the
entrance is a rate-limiting process the system can be found in
the low-density phase, while for slow exiting the high-density
phase governs the system. If the rate-limiting process is the
transition via the slowest defect bond the system is in the
maximal-current phase. In this maximal-current phase, a segregation
of density profile into macroscopic high and low regions occurs at the
location of slowest defect bond. Other defects only perturb the density
profile on a local scale. However, when the number slowest defect bonds exceeds
two or more the above picture needs modification. Furthermore, the previous studies
on disordered ASEPs lack investigations on correlation effects induced by defects.
To address these questions, we analyze the simplest model with 2 identical defects
in the bulk of the system far away from the boundaries. It was shown earlier
\cite{greulich1} that positioning of the slow defects close to the boundaries leads
only to rescaling of the effective entrance and/or exit rates, and we will not
consider this possibility in this paper. Note that in this paper we are using
terms of defect bonds and defect sites. To clarify, we define the defect site as
the site $i$ from which the particle hopes to the site $i+1$ with the rate $q<1$.
Correspondingly, the bond connecting sites $i$ and $i+1$ is a defect one.

\subsection{ Defect Mean-Field Theory}

Consider a totally asymmetric exclusion processes with open
boundaries and with 2 slow defective sites at $i=d_1$ and $i=d_2$ at a distance $d$
with $d_2-d_1=d$ (separated by $d-1$ normal sites), as shown in Fig. 1.
At the defects the particle jump to the right with the rate $q < 1$, in all
other sites the hopping rate is equal to one. It can be seen that two defects
divide the system into three segments.

\begin{figure}
\centering
\includegraphics[width=7.5cm]{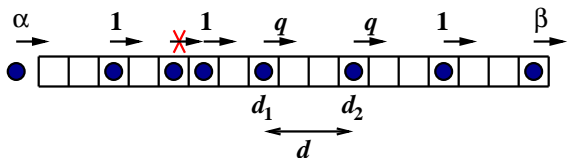}
\caption{ Fig.1: Schematic of ASEP with two defective sites at $i=d_1$ and $i=d_2$ separated by $d-1$ normal sites
i.e., $d_2-d_1=d$. The reduced hopping rates at each defect is equal to $q$.
In normal sites the hoping rates are one. } \label{fig:bz2}
\end{figure}

The particle dynamics inside each segment can be calculated exactly,
however, it is assumed that there are no correlations between the segments.
If entrance to the lattice is the slowest process then the system
can be found in low-density (LD) phase with stationary current
and bulk densities given by

\begin{equation}
J=\alpha(1-\alpha), \quad \rho_{bulk}=\alpha.
\end{equation}

Similarly, when the exit becomes a bottleneck process the system
is in high-density (HD) phase with

\begin{equation}
J=\beta(1-\beta), \quad \rho_{bulk}=1-\beta.
\end{equation}

Note that in both phases particle densities near the defect bonds
will deviate from the bulk values. The more interesting case is
when the dynamics in the system is governed by transitions via
local inhomogeneities. In this phase, which has the maximal current, we expect
to have density phase segregation similar to the case a single defect ASEPs.
We emphasize that all our investigation in this paper is on this maximal-current phase.
First consider a lattice segment after the second defect.
The dynamics in this part of the system is controlled by the entrance
of particle via the defect, then it has a low-density profile with unknown
bulk density $\rho^{*}<1/2$. Similar arguments can be used to analyze the density
profile in the segment before the first defect. Here the flux is
limited by the exit rate via the local inhomogeneity, leading to
the  high-density phase. Since at stationary-state condition
the flux through any segment should be the same,
$J=\rho^{*}(1-\rho^{*})$, the bulk density in this segment is
equal to $1-\rho^{*}$. The region between two defects can be
viewed as asymmetric exclusion process on a finite lattice with
$d$ sites. The effective entrance and exit rates to this segment
can be easily evaluated using our mean-field assumptions,

\begin{equation}\label{boundary_rates}
\alpha_{eff}=\beta_{eff}=q(1-\rho^{*}).
\end{equation}

The stationary properties of the lattice segment with $d$ sites
between the defects can be evaluated explicitly by utilizing
exact results for finite-size ASEPs \cite{derrida93}.
Specifically, the particle flux is given by

\begin{equation}\label{J0}
J_{0}(\alpha,\beta,d)=\frac{R_{d-1}(1/\beta)-R_{d-1}(1/\alpha)}{R_{d}(1/\beta)-R_{d}(1/\alpha)},
\end{equation}

where the function $R_{d}(x)$ is defined as

\begin{equation}\label{R_eq}
R_{d}(x)=\sum_{p=2}^{d+1} \frac{(p-1)(2d-p)!}{d!(d+1-p)!} x^{p}.
\end{equation}

To understand the density profile in the segment we can use a
domain-wall picture of asymmetric exclusion processes \cite{DW}.
Since the entrance and exit rates are the same [see Eq.
(\ref{boundary_rates})], the domain wall that separates
high-density and low-density blocks performs an unbiased random,
leading to a linear density profile with a positive slope.
Explicit expressions for particle densities can also be found in
Ref. \cite{derrida93}. The full description of dynamics in ASEPs
with two defects is obtained by solving for the unknown parameter
$\rho^{*}$. It can be done by applying the condition of
stationarity in the particle flux,

\begin{equation}\label{relation}
J=\rho^{*}(1-\rho^{*})=J_{0}(\alpha_{eff},\beta_{eff},d).
\end{equation}

This equation can always be solved analytically or numerically
exactly for any number of sites between local inhomogeneities,
leading to stationary particle currents and density profiles. it
is important to note that there is a particle-hole symmetry in
the system because defects are far away from the boundaries. To
illustrate our approach let us calculate dynamic properties of
ASEPs with two defects for several values of the parameter $d$.
First, let us analyze the simplest case of $d=1$ with consecutive
defects in the bulk. It can be shown that for this system

\begin{equation}
J_{0}(\alpha,\beta,d=1)=\frac{\alpha \beta}{\alpha + \beta}.
\end{equation}

Then Eq. (\ref{relation}) can be written as

\begin{equation}
\rho^{*}(1-\rho^{*})=q(1-\rho^{*})/2,
\end{equation}

which produces simple expressions for the bulk density and the
particle current,

\begin{equation}
\rho^{*}=q/2, \quad J=q(2-q)/4.
\end{equation}

The density $l$ at the site between the defects can also be found
from the condition that the flux via this site, $J=q l
(1-\rho^{*})$, should be equal to the flux through other
segments, and this yields

\begin{equation}
l=\rho^{*}/q=1/2.
\end{equation}

This result could also be obtained from the particle-hole
symmetry arguments. Note that for $q=1$ we obtain $\rho^{*}=1/2$
and  $J=1/4$ as expected for homogeneous ASEPs in the
maximal-current phase. For $d=2$ there are two lattice sites
between the defects, and stationary properties of this system can
also be obtained analytically. From Eq. (\ref{R_eq}) one can
easily derive

\begin{equation}
R_{1}(x)=x^{2}, \quad R_{2}(x)=x^{2}+x^{3},
\end{equation}

which produces the following expression for the current in the
segment between the defects

\begin{equation}\label{eq_2}
J_{0}(\alpha,\beta,d=2)=\frac{\frac{1}{\alpha}+
\frac{1}{\beta}}{\frac{1}{\alpha}+
\frac{1}{\beta}+\frac{1}{\alpha^{2}}+
\frac{1}{\beta^{2}}+\frac{1}{\alpha \beta}}.
\end{equation}

Using the expression for the effective entrance and exit rates
for the segment between the inhomogeneities [see Eq.
(\ref{boundary_rates})], the condition for the stationary current
leads to

\begin{equation}
\rho^{*}(1-\rho^{*})=\frac{2q(1-\rho^{*})}{3+2q(1-\rho^{*})}.
\end{equation}

This quadratic equation can be solved, and taking the physically
reasonable root we obtain

\begin{equation}
\rho^{*}=\frac{2q+3-\sqrt{9+12q-12q^{2}}}{4q};
\end{equation}
\begin{equation}
J=\frac{8q^{2}-6q-9 + 3\sqrt{9+12q-12q^{2}}}{8q^{2}}.
\end{equation}

It can be checked  that for $q=1$ these equations reduce to
expected relations  $\rho^{*}=1/2$ and  $J=1/4$. We can also
calculate the densities $l_{1}$ and $l_{2}$ at the sites between
the defects. Because of the particle-hole symmetry one can argue
that

\begin{equation}
l_{2}=1-l_{1},
\end{equation}

and the density at the first site  can be found by analyzing the
current via the first defect,

\begin{equation}
J=q(1-\rho^{*})(1-l_{1})=\rho^{*}(1-\rho^{*}).
\end{equation}
Then we have
\begin{equation}
l_{1}=1-\frac{\rho^{*}}{q}=\frac{4q^{2}-2q-3+\sqrt{9+12q-12q^{2}}}{4q^{2}}.
\end{equation}

We have solved equation (6) for the case $d=3$. In this case we
have:

\begin{equation}
R_{3}(x)=2x^{2}+2x^{3}+x^{4},
\end{equation}

After some lengthy but straightforward algebra we arrive at the
following cubic equation for $\rho^{*}$:

\begin{equation}
4q^2(\rho^{*})^{3} -(8q^2+6q)(\rho^{*})^{2} +
(6q^2+6q+4)\rho^{*}-q(3+2q)=0.
\end{equation}

For brevity we avoid writing the answer explicitly. Analytical
results for ASEP with 2 defects can also be obtained in the limit
of very large distances between the inhomogeneities ($d \gg 1$).
In this case the segment between the defects can be viewed as a
homogeneous ASEP in the state of the phase transition between
high-density and low-density phases ($\alpha_{eff}=\beta_{eff}$).
This corresponds to a linear density profile for the segment
between the defects. Then it leads to the following expression
for the current

\begin{equation}
\rho^{*}(1-\rho^{*})=q(1-\rho^{*})\left[1-q(1-\rho^{*})\right],
\end{equation}
and finally we obtain
\begin{equation}
\rho^{*}=\frac{q}{1+q}, \quad J=\frac{q}{(1+q)^{2}}.
\end{equation}

These results are identical to stationary properties of ASEP with
only one local inhomogeneity far away from the boundaries
obtained using DMF approximation \cite{kolomeisky98}, suggesting
that 2 defects at large distances do not affect each other
\cite{chou04}. For a general $d$ equation (6) leads to a polynomial equation
of order $d$ for the unknown $\rho^*$. For $d>3$ this equation can be solved numerically to
find the acceptable answer. In figure (2) we have sketched the behavior of current
$J$ as a function of $q$ for $d=1,2$ and $3$ and have compared them to the results
obtained via Monte Carlo simulations. As expected $J$ is an
increasing function of both $q$ and $d$. DMF notably
underestimates the current in comparison to the MC simulation
especially in the intermediate values of $q$.

\begin{figure}
\centering
\includegraphics[width=7.5cm]{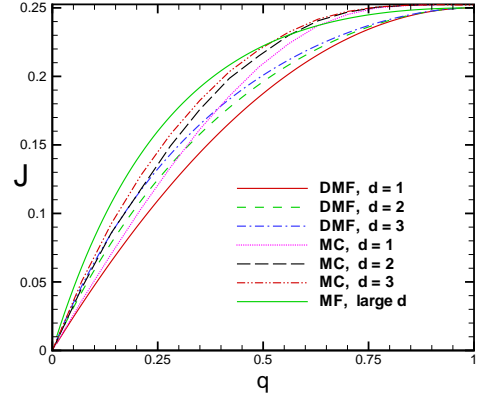}
\caption{ Fig.2: (Colour online) $J$ vs $q$ for
$d=1,2,3$ obtained by DMF method and MC simulation.} \label{fig:bz2}
\end{figure}

\subsection{Interacting Subsystem Approximation}

Interacting subsystem approximation (ISA) is another method of
calculating stationary properties of ASEPs with a single defect
or a single bottleneck developed by Greulich and Schadschneider
\cite{greulich1}.  It can be easily extended to the case of
asymmetric exclusion processes with 2 defects separated by $d$
lattice sites. Similarly to DMF this method divides the lattice
into several segments. Particle dynamics inside the segments is
treated exactly, while between the segments mean-field
assumptions are made. ISA differs from DMF in the defining of
segments. In DMF the position of defects separates different
parts, and there is always 3 segments in the system. In ISA the
sites that are connected by the defect bond are put together in
one segment, as shown in Fig. 3.

\begin{figure}
\centering
\includegraphics[width=7.5cm]{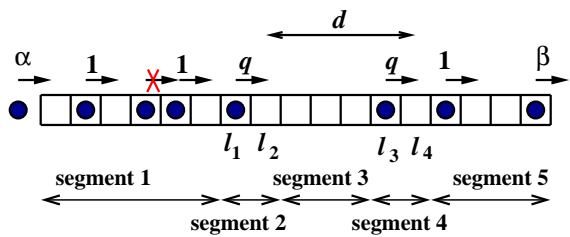}
\caption{ Fig.2: (Colour online) Fig.3: (Colour online) interacting subsystems connected via mean-field assumption.
For d$>$2 the system is divided into five segments. } \label{fig:bz2}
\end{figure}

For $d=1$ there are also 3 parts in the lattice, and the middle segment has 3 sites.
For $d=2$ there are 4 segments and 2 middle segments (with 2 lattice sites
each) border each other.  For any larger distance between local
inhomogeneities ISA assumes 5 segments: see Fig. 3. Note that the
size of the middle segment is equal to $d-2$.

Let us consider a general case of 5 segments ($d>2$) for
computation of stationary properties of ASEPs with 2 defects. As
was argued above, the system can be found in one of three phases:
LD, HD or HD/LD (maximal-current). Since the derivation of
properties in HD and LD phases is the same as for DMF approach,
we concentrate on description of the maximal-current phase. As
before we assume that the bulk density in the segments 1 and 5
are $1-\rho^{*}$ and $\rho^{*}$ correspondingly. Let us define
$l_{1}$ and $l_{2}$ as the probabilities to find the particles at
the corresponding sites of the segment around the first defect.
Similarly, $l_{3}$ and $l_{4}$ describe densities in the segment
around the second defect bond. For the middle segment with $d-2$
lattice sites we define $x_{i}$ for $i=1,\cdots,d-2$ as the
particle density at $i$-th site of this segment. As for DMF
approach, the existing particle-hole symmetry simplifies
calculations significantly. Specifically, it suggests that

\begin{equation}
l_{4}=1-l_{1},\quad l_{3}=1-l_{2}, \quad x_{i}=1-x_{d+1-i}.
\end{equation}

The overall particle current in the system can be written as

\begin{equation}
J=\rho^{*}(1-\rho^{*}),
\end{equation}

while due to the mean-field assumptions the current between the
first and the second segments is equal to

\begin{equation}
J_{12}=(1-l_{1})(1-\rho^{*})=\alpha_{2}(1-\rho^{*}),
\end{equation}

where $\alpha_{2}$ is the effective rate to enter the second
segment.  At large times we expect to find the system in the
stationary state, i.e., $J=J_{12}$, yielding

\begin{equation}
1-\alpha_{2}= l_{1}=(1-\rho^{*}).
\end{equation}

The current between segments 2 and 3 can be presented in the
several ways,

\begin{equation}
J_{23}=l_{2}(1-x_{1})=\beta_{2} l_{2}=\alpha_{3}(1-x_{1}),
\end{equation}

with $\beta_{2}$ being the effective exit rate from the segment
2, while $\alpha_{3}$ is the effective rate to enter the segment
3. When the system reaches stationary phase, $J=J_{23}$, and we
obtain

\begin{equation}
\beta_{2}=1-x_{1}, \quad
\alpha_{3}=\frac{\rho^{*}(1-\rho^{*})}{1-x_{1}}.
\end{equation}

Because of the particle-hole symmetry the effective entrance and
exit rates from the segment 3 are the same,
$\alpha_{3}=\beta_{3}$. The particle current via the ASEP segment
with $N$ sites and with entrance and exit rates $\alpha$ and
$\beta$, respectively, $J(\alpha,\beta, N)$, can be calculated
explicitly \cite{derrida93}. Then to obtain stationary properties
of ASEP  with 2 defects in the maximal-current phase the
following system of equations should be solved,

$$\rho^{*}(1-\rho^{*}) = qJ(\frac{1-\rho^{*}}{q},\frac{1-x_{1}}{q},2); $$

\begin{equation}
\rho^{*}(1-\rho^{*}) =
J(\frac{\rho^{*}(1-\rho^{*})}{1-x_{1}},\frac{\rho^{*}(1-\rho^{*})}{1-x_{1}},d-2).
\end{equation}

where $\rho^{*}$ and $x_{1}$ are 2 unknown variables. The
expression on the right side of the first equation describes the
current inside the segment 2 and 4. Because the hopping rate is
$q<1$, the effective entrance and exit rates must be rescaled by
the same factor. The right side of the second equation gives the
current inside the segment 3. The application of ISA for $d=1$
and $d=2$ cases is different. In the case of 2 consecutive defect
bonds the system is divided  only in 3 segments. The middle
segment has 3 sites that surround defect bonds. In the HD/LD
phase the effective entrance  rate is $\alpha_{2}=1-\rho^{*}$,
and the stationary properties can be obtained by solving only one
equation

\begin{equation}
\rho^{*}(1-\rho^{*})=qJ(\frac{1-\rho^{*}}{q},\frac{1-x_{1}}{q},3).
\end{equation}

Using Eqs. (\ref{J0}) and (\ref{R_eq}) for the middle segment
with equal entrance and exit rates gives us

\begin{equation}
\rho^{*}(1-\rho^{*})=\frac{q(1-\rho^{*})\left[2(1-\rho^{*})+3q\right]}{2\left[2(1-\rho^{*})^{2}+3q(1-\rho^{*})+2q^{2}\right]},
\end{equation}

which can be simplified into the following expression,
\begin{equation}
4(\rho^{*})^{3}-2(3q+4)(\rho^{*})^{2}+4(q+1)^{2}\rho^{*}-q(3q+2)=0.
\end{equation}

This cubic equation can be solved explicitly, yielding
\begin{equation}
\rho^{*}=\left[3q+4+(4-3q^{2})/D+D\right]/6,
\end{equation}
where
\begin{equation}
D=\left[-8+9q^{2}-27q^{3}+3\sqrt{3}
\sqrt{16q^{3}-q^{4}-18q^{5}+28q^{6}} \right]^{1/3}.
\end{equation}

ISA also works differently in the case of $d=2$. There are 4
segments in the system, and because of the neglect of correlations
between segments 2 and 3  we have

\begin{equation}
l_{2}(1-l_{3})=l_{2}^{2}=\rho^{*}(1-\rho^{*}).
\end{equation}

Then the effective entrance rate to the segment 2 is
$\alpha_{2}=1-\rho^{*}$, and the effective exit rate is equal to
$\beta_{2}=l_{2}=\sqrt{\rho^{*}(1-\rho^{*})}$. The unknown
parameter $\rho^{*}$ is determined from the equation for the
stationary current,

\begin{equation}
\rho^{*}(1-\rho^{*})=qJ(\frac{\alpha_{2}}{q},\frac{\beta_{2}}{q},2).
\end{equation}

Substituting the values of the effective boundary rates and
utilizing Eq. (\ref{eq_2}) we obtain

\begin{equation}
\rho^{*} \sqrt{1-\rho^{*}}+(\rho^{*})^{3/2}=q \sqrt{1-\rho^{*}}.
\end{equation}

which can be recast in the form of a cubic equation:

\begin{equation}
 2(\rho^{*})^{3} - (2q+1)(\rho^{*})^{2} + (q^2+2q)\rho^{*}-q^2=0,
\end{equation}

This equation can be solved analytically but for brevity we do not
write the solution. It can also be shown that in the limit of $d
\gg 1$ ISA method with 2 defects produces the stationary current
and bulk densities which are indistinguishable form the situation
with only one defect \cite{greulich1},
$$
\rho^{*}=\left[3q+2-\sqrt{9q^{2}-4q+4} \right]/4,
$$
\begin{equation}
J=\left[2q-9q^2+3q\sqrt{9q^{2}-4q+4} \right]/8.
\end{equation}

Let us now exhibit the dependence of $J$ on q in the ISA method.
In figure (4) we have drawn $J$ vs $q$ for $d=1,2$ and have
compared the results to those obtained by DMF method and MC
simulations.

\begin{figure}
\centering
\includegraphics[width=7.5cm]{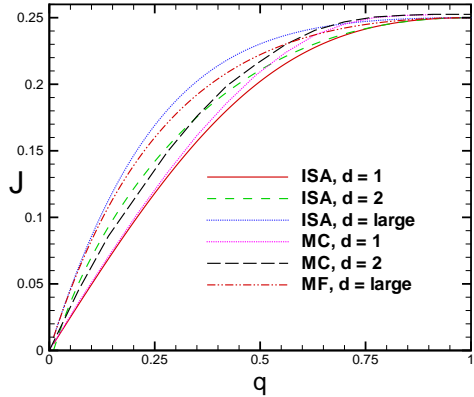}
\caption{ Fig.4: (Color online) Current vs q for
$d=1,2$ and a large $d$ obtained within ISA method and MC simulation.
In simulations we have taken $\alpha=\beta=0.6$. } \label{fig:bz2}
\end{figure}

In general, ISA method gives a better estimation of current
compared to DMF at least for small values of $d$ we have
considered.

\section{Correlations near defect}

\subsection{Monte Carlo Simulations}

In this section we aim to investigate correlations in the
vicinity of defects. We restrict ourselves to adjacent two-point
correlations and will present our results for the general two-point and
multi-point correlations in a future work. Let us now introduce the
normalized connected two-point correlation function $C_i$ between the neighbouring
sites $i$ and $i+1$. This quantity is defined as follows:

\begin{eqnarray}
C_i= \frac{\langle \tau_i\tau_{i+1} \rangle - \langle \tau_i
\rangle \langle \tau_{i+1} \rangle}{ \sqrt{\langle \tau_i^2
\rangle- \langle \tau_{i} \rangle ^2}\sqrt{\langle \tau_{i+1}^2
\rangle- \langle \tau_{i+1} \rangle ^2}} ~ i=1,\cdots,L-1.
\end{eqnarray}

The function $C_i$ measures the correlation and it lies between
$-1$ and $1$. Negative values correspond to anti-correlation
between neighboring sites whereas a positive value signifies
correlation. The values near zero are regarded as uncorrelated.
Fig. (5) depicts the simulated profiles of correlation at $d=10$
and 100 each for three values of $q$. The system size is $L=500$ and we have
taken $\alpha=\beta=0.6$ in all our simulation results unless stated otherwise.
The system has been updated for $T$ Monte Carlo steps. Each step consists of $L$ moves.
In each move, we randomly choose a site and update its status according
to ASEP rules described above. We discard the first $\frac{T}{5}$ steps to
ensure reaching steady state, and we have accumulated data
separated by 10 MC steps to avoid any possible temporal
correlations. The value of $T$ is taken $10^8$ in our simulations.

\begin{figure}
\centering
\includegraphics[width=7cm]{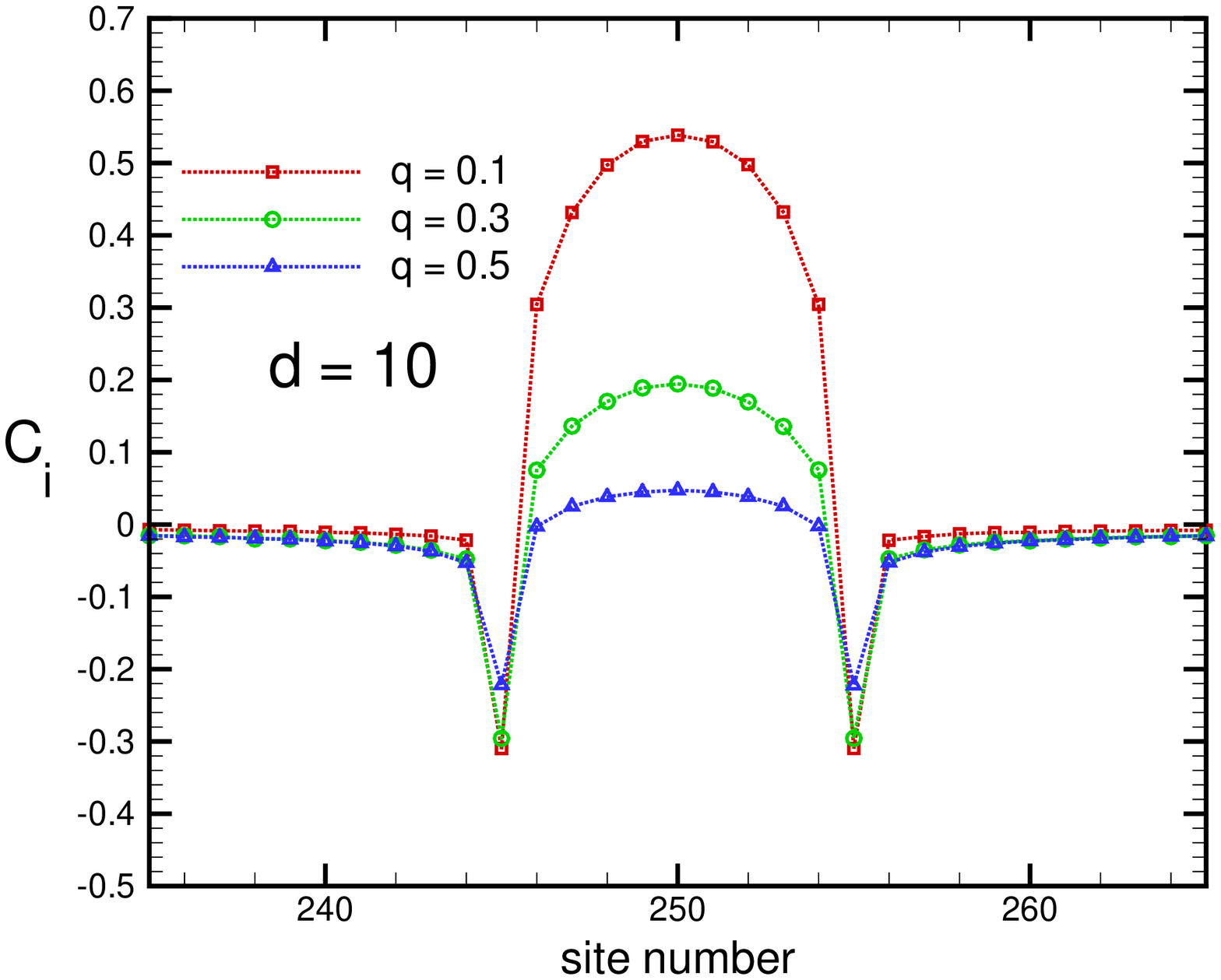}
\includegraphics[width=7cm]{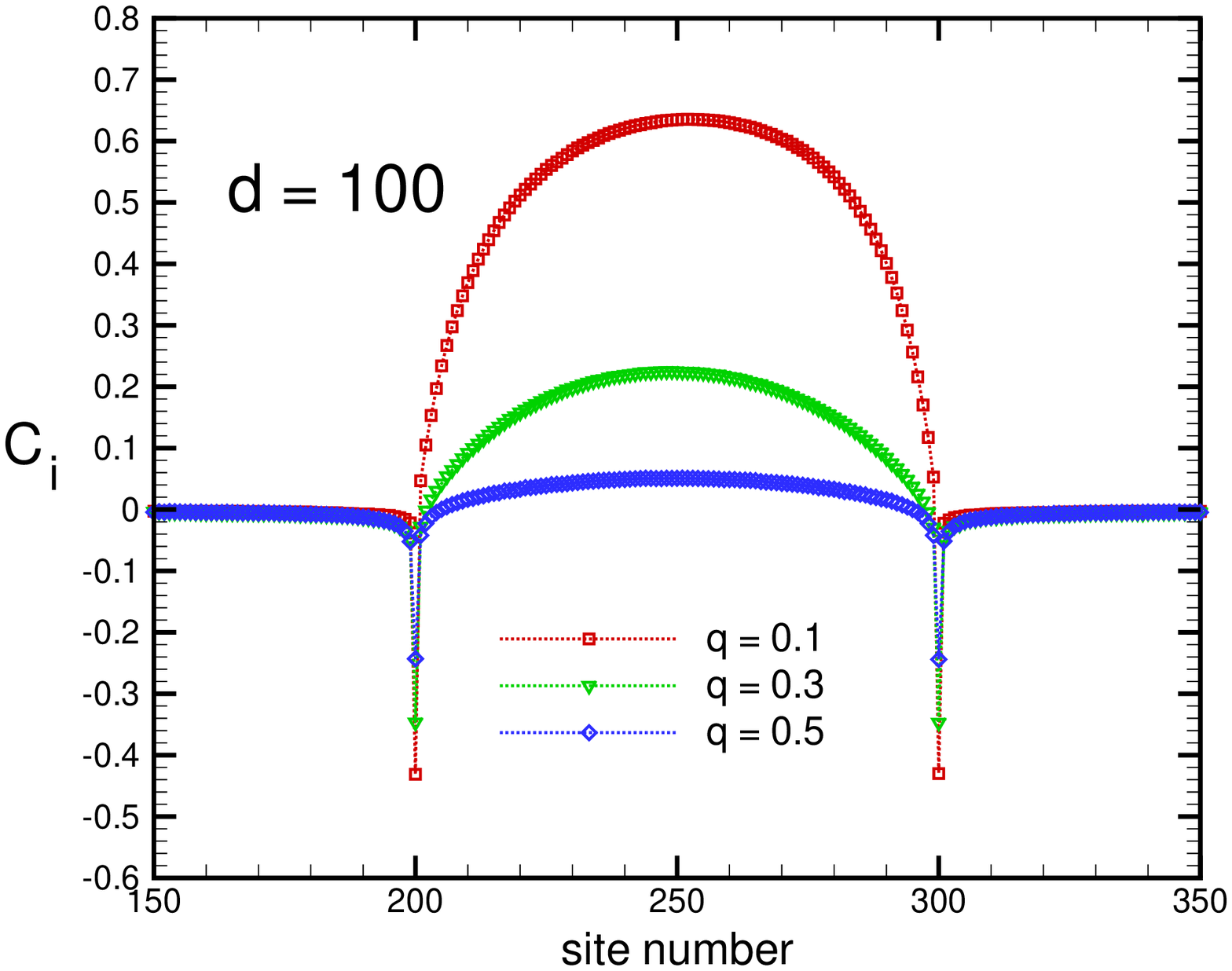}
\caption{ Fig.5: (Color online) Profile of normalized correlation function at $d=10$ (top)
and $d=100$ (bottom) for $q=0.1,0.3,0.5$. } \label{fig:bz2}
\end{figure}

Two defects are symmetrically placed with respect to chain mid
point. We observed that correlations are large in sites
between the defects. There is a rather strong anti-correlation in
the sites immediately after the first defect and before the
second defect. The correlations are growing up for middle sites
where the maximum value is achieved. It can be seen that
correlations are greater when $d$ is increased. This is unexpected
and counterintuitive because increasing the distance between the defects
reduces their interaction. It has been observed via MC
simulations that when $d$ is increased the current reaches
asymptotically to its mean-field value
$J_{MF}=\frac{q}{(1+q)^2}$  \cite{dong07,dongPRE}. Therefore, one
expects the correlations to exhibit a reducing behavior with
respect to distance $d$ but this is not observed in our
simulations. To have a deeper understanding, we have sketched the
behavior of correlation profile upon varying the distance $d$ for
$q=0.1$ and $q=0.3$ in Fig. 6.

\begin{figure}
\centering
\includegraphics[width=7cm]{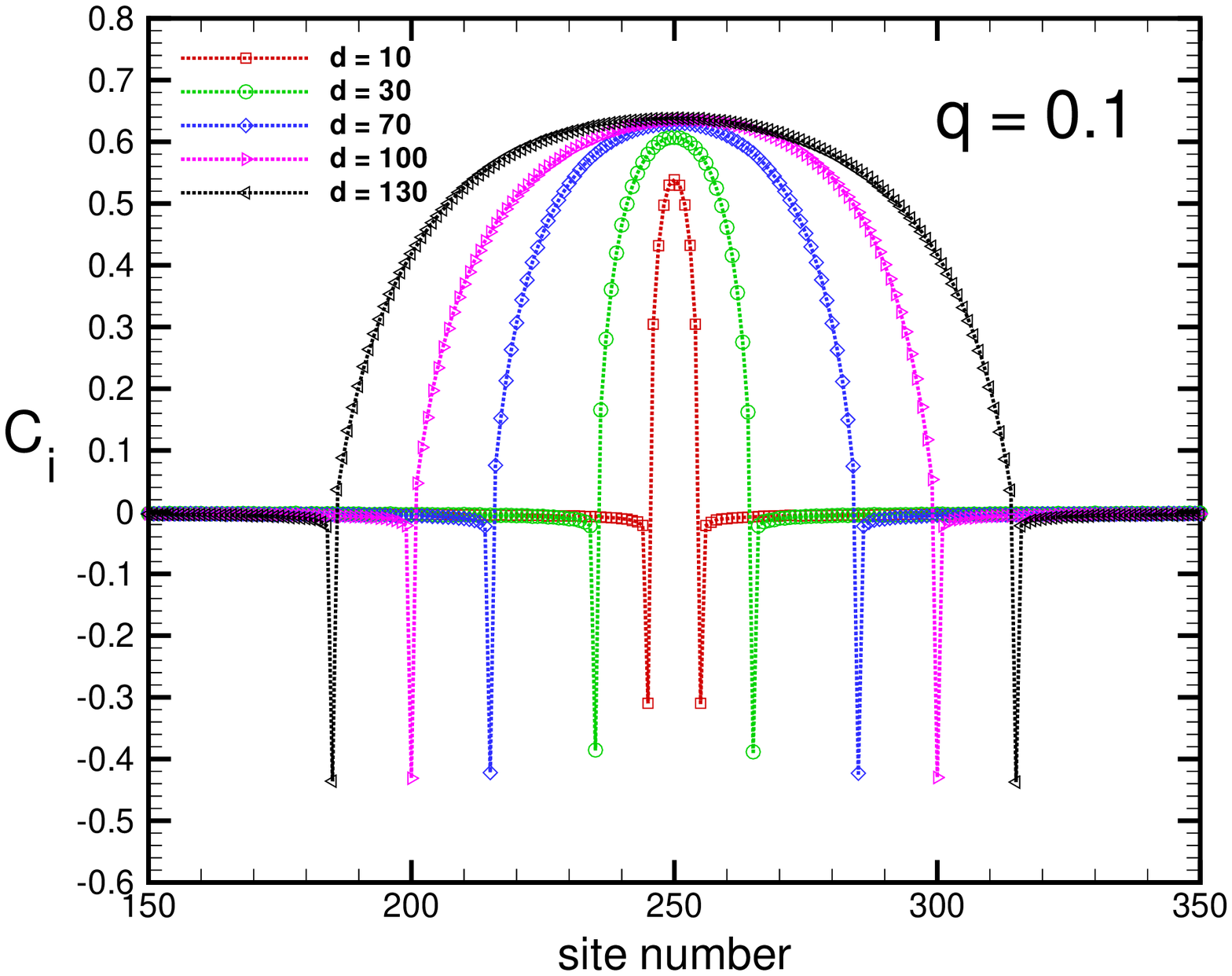}
\includegraphics[width=7cm]{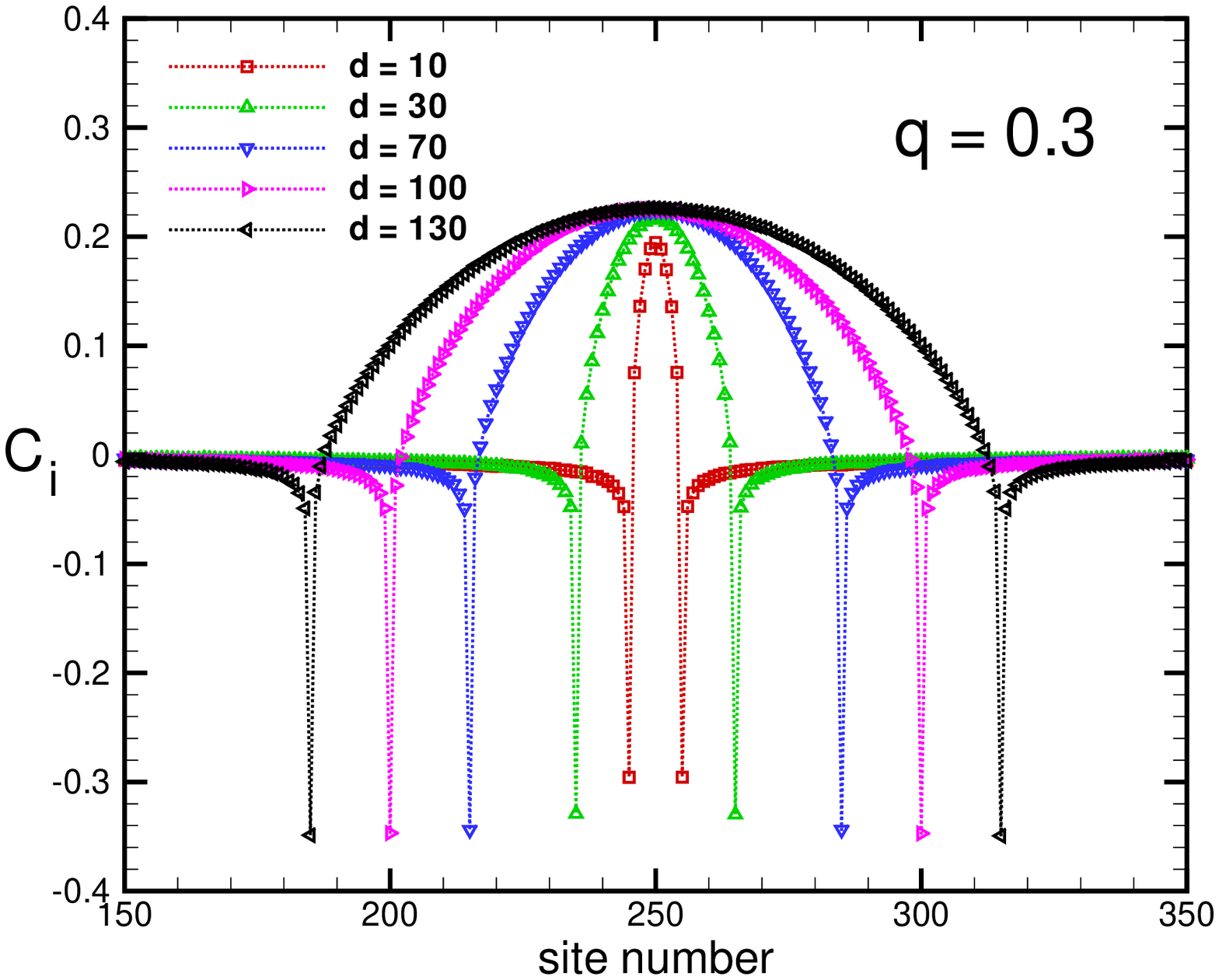}
\caption{ Fig.6: (Color online) Profile of
normalized correlation functions at various values of $d$ for
$q=0.1$ (top) and $q=0.3$ (bottom). } \label{fig:bz2}
\end{figure}

For fixed values of $q$, increasing the distance $d$ between the
defects gives rise to enhancement of
correlations/antocorrelations. For instance, the correlation value
in the middle point rises from roughly 0.5 at small $d \sim 10$
to 0.65 for $d \sim 100$. It can be observed that there is no
notable difference in correlation values for $d$ larger than
$100$. Moreover, the correlations are always greater than
anti-correlations. By increasing $q$, the correlations/anti-correlations
are notably reduced in values. This is expected since in the limit of
homogeneous ASEP where $q\rightarrow 1$ the correlation functions become very small.
Here we wish to make a pause and have a discussion on correlations in normal ASEPs.
In fact the middle segment between two defects can be regarded as an ASEP chain with
length $d$ with equal entrance and exit rates. To the best of our knowledge, correlations
in ASEP with random sequential update, has only been analytically discussed by
Derrida and Evans who obtained exact analytical expression for a general two-point function
and made a conjecture to generalize their findings to $n$-point function \cite{evans93}. Their
study was restricted to the special case $\alpha=\beta=1$ and they found that
long range correlations persist in the bulk which was attributed as a boundary effect.
In order to see if the large value of the connected two-point function survives in
the normal ASEP with equal entrance and exit rates, we performed MC simulations. Our
results show that when $\alpha=\beta$, the profile of $C_i$ reaches a small constant (almost zero) in the bulk.
The correlations become large near boundaries. This boundary behaviour depends on whether $\alpha=\beta <0.5$
or $\alpha=\beta >0.5$. Figure (7) illustrates this aspect.

\begin{figure}
\centering
\includegraphics[width=7cm]{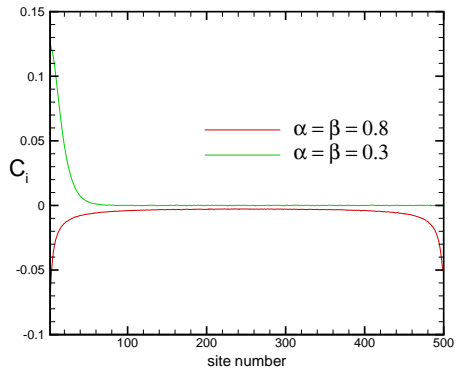}
\caption{ Fig.7: (Color online) Profile of normalized correlation functions in a normal ASEP chain with $\alpha=\beta$ .
} \label{fig:bz2}
\end{figure}

We recall that correlations in other types of update such as parallel updating has been discussed in
\cite{schutz93,schutz93PRE}. It is worthwhile to examine the behavior of density profile between
defects. Dong et al have recently shown via extensive MC simulations that
the density profile takes a linear shape between defects \cite{dong07}. This behavior remains
unchanged in ASEP with extended objects \cite{dongPRE}. For the
sake of completeness, we show some typical density profiles in Fig. 8.

\begin{figure}
\centering
\includegraphics[width=7cm]{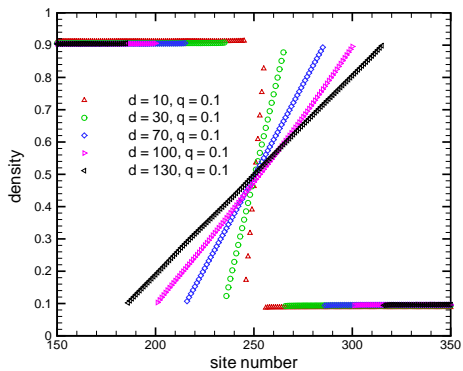}
\caption{ Fig.8: (Color online) Profile of density at various values of $d$ for $q=0.1$.
The profile exhibits a linear behaviour with positive slope. This behaviour is associated
to the existence of wandering shock in the region between left and right defects. } \label{fig:bz2}
\end{figure}

The interesting point is the absence of boundary layer in this
phase-segregated regime. It would be illustrative
to look at the dependence of two-point correlation
functions at some particular sites on values of $q$ and $d$.
These results are sketched in Fig. 9 where correlations at the
first defect site ($d_1$), its rightmost site ($d_1+1$) and in
the middle site of the chain are considered.

\begin{figure}
\centering
\includegraphics[width=7cm]{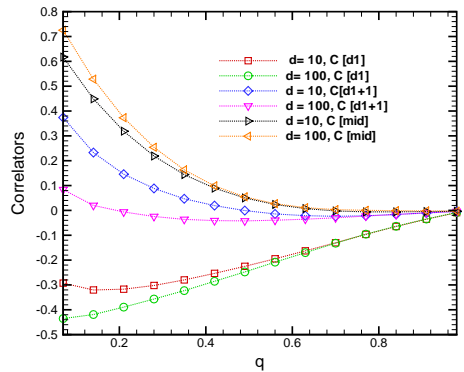}
\includegraphics[width=7cm]{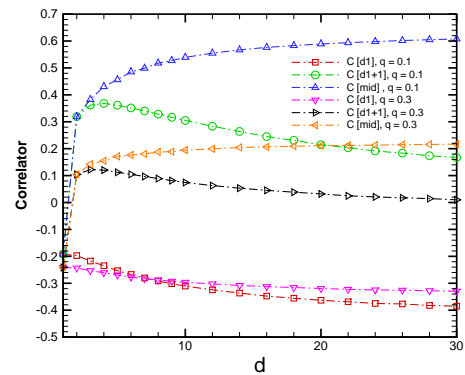}
\caption{ Fig.9: (Color online) Dependence of normalized correlation functions at three
selected sites on $q$ (top) and on $d$ (bottom). } \label{fig:bz2}
\end{figure}

Note that all correlations/anti-correlations approach zero when
$q$ tends to one. Moreover, increasing defect's separation $d$
increases the correlations. The dependence on $d$ is more
interesting. While the values of correlation functions reach an
asymptotic value at large d, the behavior is not monotonous.
Especially for $C_{d_1+1}$ correlation increases up to a maximum
and then begin to decrease smoothly towards its asymptotic value.
The value of $d$ where $C_{d_1+1}$ is maximized does not show a
significant dependence on $q$. In can be concluded that varying
$d$ can dramatically affect the system characteristics as far as
correlations are considered.

\subsection{Analytical theory }

Our simulation findings in the preceding section confirms that in
between the defects the correlations are notably higher than other
sites. In this section we try to develop a theoretical framework
to capture this feature. Suppose we have two slow defects located
in the bulk at sites $k$ and $l$ ($k<l$) respectively both with rate
$q$. The mean occupation at
site $i$ in the steady state is denoted by $n_i=\langle \tau_i
\rangle$, in which $\tau_i=0,1$ is the occupation number at site
$i$. We assume that a simple MF assumption, i.e., $\langle
\tau_i\tau_{i+1}\rangle =\langle \tau_i \rangle \langle
\tau_{i+1}\rangle =n_in_{i+1}$ holds for all sites except
$i=k,\cdots,l$ i.e.; defective sites themselves and all the sites between them.
At these sites the correlations are strong enough to violate the simple
mean-field assumption. Let us introduce two-point correlation
functions $m_i$ in the following way,

\begin{eqnarray}
m_i=\langle \tau_{i}\tau_{i+1}\rangle.
\end{eqnarray}

There are $L+l-k+2$ unknowns, namely, $
n_1,n_2,\cdots,n_L,m_{k},m_{k+1},\cdots,m_{l}$ and lastly the
current $J$. In the stationary state there exists $L+1$ equations among these unknowns.
Let us label them by $A_0$ to $A_{L}$. These
equations can be obtained by expressing the current $J$ in terms
of the mean site densities and two point correlators. The first
equation, $A_0$, is $J=\alpha(1-n_1)$. The equations $A_i
$ for $~i=1,\cdots,k-1$ and $i=l+1,\cdots,L-1$ have the following form,

\begin{eqnarray}
J= n_i(1-n_{i+1}).
\end{eqnarray}

Equation $A_k$ is:
\begin{eqnarray}
J=q \langle \tau_k(1-\tau_{k+1}) \rangle =q(n_k-m_k).
\end{eqnarray}
Similarly, equation $A_l$ is given below
\begin{eqnarray}
J=q \langle \tau_{l}(1-\tau_{l+1}) \rangle =q(n_l-m_l).
\end{eqnarray}

Equations $A_i~ (i=k+1,\cdots,l-1) $ have the following form

\begin{eqnarray}
J= \langle \tau_{i}(1-\tau_{i+1}) \rangle =(n_i-m_{i+1}).
\end{eqnarray}

lastly equation $A_L$ is as follows,

\begin{eqnarray}
J=\beta \langle \tau_L\rangle=\beta n_L.
\end{eqnarray}

We do not intend to add more equations. Unfortunately the above
equations are nonlinear and it would be a formidable task to solve
them analytically. Alternatively, we shall utilize a numerical
approach to solve the system of equations by exploiting their recursive structure.
This approach was originally introduced in \cite{foulaadvand07} in the context of
disordered ASEP and was later applied to the problem of two
intersecting ASEP chains \cite{foulad07}. According to this
numerical scheme, we assign a value to $J$. Having $J$, it is possible
to iterate equations (in forward direction) and obtain $n_1$ up to $n_k$. Then we
proceed to find $m_k$ by incorporating the relation $J=q(n_k
-m_k)$. At this stage it is not possible to proceed further because both
$n_{k+1}$ and $m_{k+1}$ are unknown and we have only one relation
between them : $J=n_{k+1} -m_{k+1}$. In order to proceed, we
approximate $n_{k+1}$ in the following way. Consider the rate equation for
the 2-point function $\langle \tau_{k} \tau_{k+1} \rangle$ which is governed by
the following master equation:

\begin{eqnarray}
\frac{d\langle \tau_{k} \tau_{k+1} \rangle }{dt}= \langle
\tau_{k-1}(1-\tau_k)\tau_{k+1}\rangle - \langle
\tau_{k}\tau_{k+1}(1-\tau_{k+2})\rangle.
\end{eqnarray}

In the steady state, the left hand side becomes zero, and
therefore two terms on the right hand side will be equal. To
proceed further we have to approximate 3-point functions. This is
achieved by utilizing the cluster mean-field assumption \cite{chowdhury02}.
According to this assumption we replace any three point function
by the product of 2-point functions as follows,

\begin{eqnarray}
\langle n_in_jn_k\rangle=\frac{\langle n_in_j\rangle\langle
n_jn_k\rangle}{\langle n_j\rangle}.
\end{eqnarray}

We then replace all 2-point functions by the product of 1-point
functions except $m_k=\langle \tau_{k}\tau_{k+1} \rangle$.
Then it is possible to express $1-n_{k+2}$ in
terms of $n_{k-1},m_1$ and $n_{k+1}$. After some algebra a quadratic
equation for $n_{k+1}$ is obtained:

\begin{eqnarray} n_{k-1}n_{k+1}^2 - m_kn_{k-1}n_{k+1}
-m_k J=0.
\end{eqnarray}

The physically reasonable solution for this equation is given by

\begin{eqnarray}
n_{k+1}= \frac{ m_k  + \sqrt{ m_k^2 + \frac{4m_kJ}{n_{k-1}} }
}{2}.
\end{eqnarray}

Now it is possible to find $m_{k+1}$ via equation $J=n_{k+1} -m_{k+1}$. Analogous to the above
procedure we can find $n_{k+2}$ as follows:

\begin{eqnarray}
n_{k+2}= \frac{ qm_km_{k+1}  + \sqrt{ (qm_km_{k+1})^2 + 4qJn_kn_{k+1}^2m_{k+1} }
}{2qn_kn_{k+1}}.
\end{eqnarray}

After having $n_{k+2}$ we simply obtain $m_{k+2}$ via equation $J=n_{k+2} -m_{k+2}$. Now it would be possible
to proceed iteratively after taking into account some approximation. To this end we recall the equality

\begin{eqnarray}
\frac{d\langle \tau_{i} \tau_{i+1} \rangle }{dt}= \langle
\tau_{i-1}(1-\tau_i)\tau_{i+1}\rangle - \langle
\tau_{i}\tau_{i+1}(1-\tau_{i+2})\rangle.
\end{eqnarray}

Putting the left hand side equal to zero in the steady state, utilizing cluster mean-field in three point functions and finally
substituting $m_{i+1}$ by $m_{i+1}=n_{i+1} -J$ we arrive at the following equation:

\begin{eqnarray}
n_{i+1}= \frac{ m_{i-1}m_{i}  + \sqrt{ (m_{i-1}m_{i})^2 + 4Jn_{i-1}n_{i}^2m_{i} }
}{2n_{i-1}n_{i}}.
\end{eqnarray}

Note that we have approximated $\langle \tau_{i-1} \tau_{i+1} \rangle$ by the mean-field relation $\langle \tau_{i-1} \rangle \langle \tau_{i+1} \rangle$.
We can iterate equation (53) together with $m_{i+1}=n_{i+1}-J$ from $i=k+2$ to $l-2$ to find the corresponding $n_i$ and $m_i$ up to $i=l-1$.
The site $i=l$ needs to be treated separately. Following the same strategy we easily find:

\begin{eqnarray}
n_{l}= \frac{ m_{l-2}m_{l-1}  + \sqrt{ (m_{l-2}m_{l-1})^2 + 4Jn_{l-2}n_{l-1}^2m_{l-1} }
}{2n_{l-2}n_{l-1}}.
\end{eqnarray}

From which one can compute $m_l$. In a similar fashion, we can obtain $n_{l+1}$. We only should
shift up all the subscripts in equation (54) by one. Now it is possible again to proceed iteratively to the end of the chain and
evaluate $n_L$ which gives us the output current $J^{out}$. If the given value of input $J$ were correct,
the output current $J^{out}$, which is $\beta n_L$, should be the same, up to a given precision, as the input value of $J$.
By systematically increasing the input $J$ in an small amount $\delta J$, we can determine the correct $J$ and
correspondingly the mean densities $n_1,\cdots,n_L$ together with correlators $m_k,\cdots,m_l$.
In Fig. (10) the dependence $J$ on $q$ obtained by the numeric scheme devised above
is sketched. For the sake of comparison, we have augmented the
figure with the analogous graphs obtained by MF, DMF, ISA and MC
methods.

\begin{figure}
\centering
\includegraphics[width=7cm]{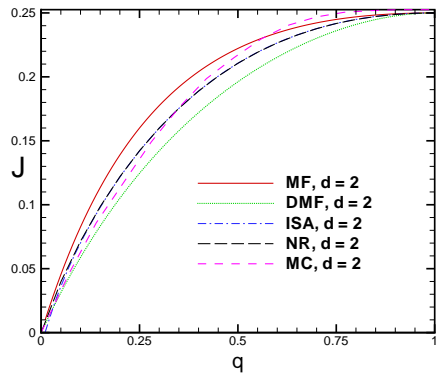}
\caption{ Fig.10: (Color online) Current vs q for $d=2$ obtained by various analytical and
numerical methods and MC simulation. $J$ approaches to 0.25 when $q$ tends to one in accordance to normal ASEP.
} \label{fig:bz2}
\end{figure}

The result of the numerical scheme is almost identical to ISA
method. They both are in very good agreement with Monte Carlo
simulations. However, the numerical scheme has an advantage over
ISA method in the sense that it can easily be implemented for any
$d$, whereas finding the solution of the ISA nonlinear set of
equations, i.e., Eq. (29) is not an easy task even by employing
advanced numerical methods. Finally in figure (11) we have sketched the dependence of $J$ on $d$ obtained
from MC simulation and the numeric scheme.

\begin{figure}
\centering
\includegraphics[width=7cm]{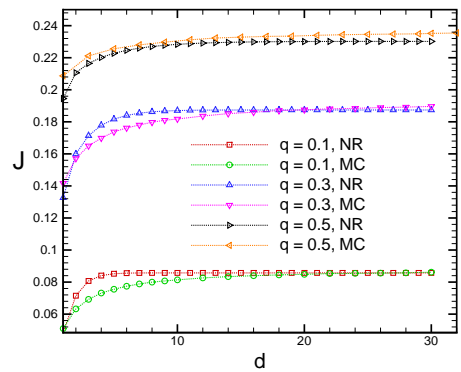}
\caption{ Fig.11: (Color online) Current vs $d$ for $q=0.1,0.3$ and $0.5$ obtained by the
numeric method and MC simulation. } \label{fig:bz2}
\end{figure}

The results of the numeric method are in rather good agreement
with those obtained by MC simulations. $J$ is an increasing
function of $d$ and becomes saturated after some short
$q$-dependent distance. The results confirms the earlier finding
in \cite{dong07}. Note that the length scale on which $J$
recovers its single-defect value is of the same order of
magnitude of the correlation length in the density profile near
boundaries. Finally we would like to add that our numerical scheme
is not capable of reproducing the profile of correlators obtained
via MC simulations. Figure (12) depicts the profile of
unnormalised adjacent two-point correlation function $\langle
\tau_{i+1}\tau_i \rangle - \langle \tau_{i+1} \rangle \langle \tau_i
\rangle$ for two methods of simulation and numerical scheme.

\begin{figure}
\centering
\includegraphics[width=7cm]{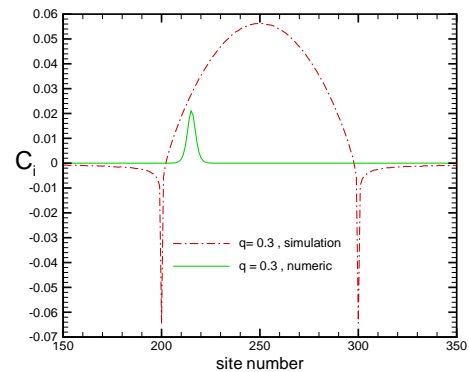}
\caption{ Fig.12: (Color online) Profile of corrrelators  for $d=100$ and $q=0.3$. } \label{fig:bz2}
\end{figure}

We see that within the numerical framework both the value and the extension of correlators are small in comparison to
the simulation results. The reason lies in the implementation of a series of approximation in several places in this
numerical algorithm.

\section{Summary and Conclusions}

An open ASEP chain with two defective sites with reduced hopping rates $q<1$ has been investigated. The system current and mean site densities
at defective sites and their vicinities have been obtained by two analytical methods namely defect mean-field (DMF) and interacting
subsystem approximation (ISA). Both methods combine mean-field approach near defects with known exact solutions. Our results are accomplished
by extensive Monte Carlo simulations. We focus on the phase-segregated phase in which defects globally affect the system properties and
the system is not input/output rate-limited but rather defect-limited. MC simulations have revealed strong short ranged correlations at the
defective sites and at all the site between them. Additionally, the profile of density between defects takes a linear form which marks the
existence wandering shock in this intermediate region. In order to take into account these correlations, we have introduced a numerical approach
which utilizes a cluster mean-field assumption. Comparison of the three methods show that ISA and the numeric approach give a current value which
is in a good agreement with MC simulations. DMF method however, only gives a good results compared to MC and other two methods for small $q$ less
than $0.1$ which is due to strong correlations. Furthermore, we have obtained the profile of neighbouring two-point correlation function throughout
the chain. It is shown that these two point correlators exhibit a rather strong anti correlation at the first defective site then they grow to the
middle of the defects and after that they start diminishing. In general, two point correlators will tend to a tiny value when $q$ approaches one.
However, the dependence of two point correlators for a fixed $q$ as a function of distance between defects is not monotonous. Despite reaching
to an asymptotic value for large distances, one observes a peak at short distances for the two point correlators near defects. Our theoretical
analysis indicates that correlations are critically important for dynamics of particles in disordered ASEPs. It also shows that it is possible
to devise an approximate method that can take into account these correlations, providing a satisfactory description of stationary properties.

\section*{Acknowledgments}

ABK acknowledges the support from the Welch Foundation (under Grant
No. C-1559), and from the US National Science Foundation (grants
CHE-0237105 and NIRT ECCS-0708765). MEF expresses his gratitude to
M. F. miri for their useful help.

\end{document}